\documentclass[paper]{JHEP3}
\usepackage{latexsym,amssymb}
\usepackage{amsmath}
\usepackage{amscd}
\usepackage{amsfonts}
\usepackage{amsthm}
\usepackage{graphicx}
\allowdisplaybreaks[1]

\def\half{{\textstyle {1 \over 2}}}
\def\quart{{\textstyle {1 \over 4}}}
\def\noverm#1#2{{\textstyle {#1 \over #2}}}

\def\sgn{{\rm sgn}}
\def\tr{\,{\rm tr}\,}
\def\makeatletter{\catcode`\@=11}
\makeatletter
\def\mathbox#1{\hbox{$\m@th#1$}}%
\def\math@ccstyles#1#2#3#4#5#6#7{{\leavevmode
      \setbox0\mathbox{#6#7}%
      \setbox2\mathbox{#4#5}%
      \dimen@ #3%
      \baselineskip\z@\lineskiplimit#1\lineskip\z@
      \vbox{\ialign{##\crcr
             \hfil \kern #2\box2 \hfil\crcr
             \noalign{\kern\dimen@}%
             \hfil\box0\hfil\crcr}}}}
\def\mathaccstyles{\math@ccstyles\maxdimen}
\def\maththroughstyles{\math@ccstyles{-\maxdimen}}
\def\unitmatrixDT%
 {\maththroughstyles{.45\ht0}\z@\displaystyle {\mathchar"006C}\displaystyle 1}
\def\leftrightarrowfill{$\mathsurround0pt \mathord\leftarrow
       \mkern-6mu\cleaders\hbox{$\mkern-2mu \mathord- \mkern-2mu$}\hfill
       \mkern-6mu \mathord\rightarrow$}

\def\overleftrightarrow#1{\vbox{\ialign{##\crcr
        \leftrightarrowfill\crcr\noalign{\kern-1pt\nointerlineskip}%
        $\hfil\displaystyle{#1}\hfil$\crcr}}}

\def\unitm#1{\unitmatrixDT_{#1}}

\title{Potential and mass-matrix in gauged $N=4$ supergravity}

\author{M.~de Roo and D.~B.~Westra,\\
Institute for Theoretical Physics\\
   Nijenborgh 4, 9747 AG Groningen,
     The Netherlands\\
     E-mail: \email{m.de.roo@phys.rug.nl,\ d.b.westra@phys.rug.nl}}
\author{S.~Panda\\
    Harish-Chandra Research Institute \\
    Chatnag Road, Jhusi, Allahabad 211019, India\\
        E-mail: \email{panda@mri.ernet.in}}
\author{M.~Trigiante\\
    Spinoza Institute\\
    Leuvenlaan 4, 3508 TD Utrecht,
    The Netherlands\\
    E-mail: \email{m.trigiante@phys.uu.nl}}
\preprint{UG-03/06\\ MRI-P-031001\\SPIN-03/36\\ITP-UU-03/55 \\ \hepth{0310187}}

\abstract{We discuss the potential and mass-matrix of gauged
$N=4$ matter coupled supergravity for the case of six matter multiplets,
extending previous work by considering the dependence on all
scalars. We consider all semi-simple gauge groups
and analyse the potential and its first and second derivatives
in the origin of the scalar manifold.
Although we find in a number of cases an extremum with a positive
cosmological constant, these are not stable under
fluctuations of all scalar fields.
}


\begin{document}

\section{Introduction\label{Intro}}

Recent astronomical observations \cite{WMAP,SN1A} have led to the
conclusion that our universe is presently in a state of
accelerating expansion, and that its cosmological properties are
best described by assuming the presence of a large amount of dark
matter of unknown origin, as well as an even larger amount of dark
energy, which one considers as due to a cosmological constant.
This situation has renewed interest in fundamental theories with
scalar potentials, as these may have nonzero extremal values, thus
presenting a possibility to explain the cosmological constant.
Motivated by string and M theory, the search for the appropriate
fundamental theory concentrates on the supergravity theories which
arise as low energy limit of compactified string theories. A major
problem in this respect is that the sign of the cosmological
constant should be positive to explain the expansion, a property
which is hard to reconcile with theories which have their origin
in string theory (\cite{GWG,dWHDS,MN}, for a recent discussion see
\cite{PT}).

In this paper we take the point of view that one should first obtain
a solution of a four-dimensional supergravity theory with the appropriate
properties (de Sitter, stability), and consider the connection with string
theory as a second step.
We will concentrate on the properties of gauged $N=4$ supergravity
theories coupled to additional matter multiplets. It is well-known that
these theories do allow extrema with a positive cosmological constant
\cite{GZ}. In a
previous paper \cite{dRWP1} a situation where the scalar manifold is truncated
to four scalar fields was discussed, and it was shown that a positive,
stable extremum
of the potential is possible when the scalar fluctuations are limited
to the four directions which survive the truncation. In the present  paper we will
extend the scope of these investigations to include more general gauge groups,
and also  consider the fluctuations in all scalars present in
the model.

A further motivation for this investigation arises from  \cite{FTVP},
where a stable solution in de Sitter space was constructed
in gauged $N=2$ supergravity. There is no obvious connection of this solution
with string theory, but the authors indicated a possible
relation with $N=4$. Since $N=4$ is a step closer to the maximal ten and eleven
dimensional supergravity theories, the connection with string theory might
be easier to obtain once the $N=2$ case is raised to $N=4$. In Section
\ref{Conc} we will discuss the $N=2$ aspect of our work in more detail.

We leave for later work the question of how these theories arise
from the higher dimensional low-energy limit of string theory. It
is well known that ungauged $N=4$ supergravity theories in $d=4$
can be obtained by toroidal compactification from $d=10$
supergravity \cite{ACNP81}. Gauged $N=4$ supergravity obtained
from Scherk-Schwarz compactifications \cite{SS} has been
considered in the past \cite{ACPR81,TW84,PZ89}, and also more
recently in combination with flux compactification
\cite{BdRE,AFV,ADFL,AFLV,D'Auria:2003jk,AFT}. Nevertheless, to our knowledge the
$SU(1,1)$ duality angles \cite{MdRPW1}, which play
an essential role in obtaining extrema with a positive
cosmological constant, have not been given a higher-dimensional
origin.

Throughout this paper we will use the notation of and results from \cite{dRWP1}.
We will discuss the value of the potential and its derivatives, as well as
the definition of the bosonic mass-matrices, in Section \ref{Pot}.
Nine semi-simple gauge groups
satisfy the conditions that (i) the potential has an extremum
for all scalar fields, (ii) the value of the potential in the extremum is
positive. These groups are introduced in Section \ref{firstder}
and their properties are discussed in detail in the different subsections
of Section \ref{groups}.

To complete this Introduction we will review some basic
properties of the parametrisation of the
scalar sector of matter coupled $N=4$ supergravity \cite{MdR1,MdRPW1}.
We consider gauged $N=4$ supergravity coupled to $n$ vector multiplets.
The scalar fields of the theory are\footnote{The
indices $\alpha,\beta,\ldots$ take on values $1$ and $2$,
indices $R,S\ldots$ the values $1,\ldots, 6+n$, and
the indices $a,b,\ldots$ the values $1,\ldots,6$. The metric $\eta_{RS}$ can
be chosen as diag$(-1,-1,-1,-1,-1,-1,+1,\ldots,+1)$, with $n$ positive
entries. In comparison to \cite{MdRPW1} we have replaced
the complex scalars $\phi_{ij}{}^R$ by real scalars $Z_a{}^R$:
$\phi_{ij}{}^R = \half Z_a{}^R (G^a)_{ij}$, where the $G^a$ are
six matrices which ensure that $Z_a{}^R$ transforms as a
vector under $SO(6)$.}
$Z_a{}^R$ (real) and $\phi_\alpha$ (complex), satisfying the constraints
\begin{eqnarray}
\label{conphi}
   &&\phi^\alpha\phi_\alpha = 1\,,
   \\
\label{conZ}
   &&\eta_{RS}Z_a{}^R Z_b{}^S = -\delta_{ab}\,.
\end{eqnarray}
The scalars $\phi_\alpha$ ($\phi^1=(\phi_1)^*, \phi^2=-(\phi_2)^*$)
transform under global $SU(1,1)$ and
local $U(1)$, the $Z_a{}^R$ transform under local $SO(6)\times SO(n)$, and
under global $SO(6,n)$. The constraints and the local symmetry
restrict the scalars to the
cosets $SU(1,1)/U(1)$ (two physical scalars) and $SO(6,n)/SO(6)\times SO(n)$
($6n$ physical scalars).

There is a certain freedom in coupling the vector multiplets: for each
multiplet, labeled by $R$, we can introduce an $SU(1,1)$ element, of which only a
single angle $\alpha_R$ turns out to be important. These angles $\alpha_R$
can be reinterpreted as a modification of the $SU(1,1)$ scalars coupling to
the multiplet $R$ in the form
\begin{equation}
    \phi^1_{(R)} = e^{i\alpha_R}\phi^1\,,\quad
    \phi^2_{(R)} = e^{-i\alpha_R}\phi^2\,,\quad
    \Phi_{(R)} = e^{i\alpha_R}\phi^1 + e^{-i\alpha_R}\phi^2\,.
\end{equation}
The gauge group has to be a subgroup of $SO(6,n)$.
For a semi-simple gauge group the $\alpha_R$ (called
$SU(1,1)$ angles in this paper) have to be the same for all
$R$ belonging to the same factor of the gauge group. This gauging
breaks the global $SO(6,n)$ symmetry of the ungauged theory.
In the remainder of this paper we will limit ourselves to $n=6$.
The reason for the choice $n=6$ is, besides its relative simplicity,
that this case follows by toroidal compactification from $d=10$
$N=1$ supergravity without additional matter.

\section{The potential, its derivatives, and the mass-matrix\label{Pot}}

\subsection{The potential\label{potential}}

The scalar potential $V(\phi,Z)$ can be written in the form
\begin{equation}
\label{pot}
   V = \sum_{i,j}\, (R^{(ij)}(\phi)\, V_{ij}(Z) + I^{(ij)} \,W_{ij}(Z))\,.
\end{equation}
The indices $i,j,\ldots$ label the different factors in the gauge group
$G$, which we will take to be semi-simple.
$R^{(ij)}$ and $I^{(ij)}$ contain the $SU(1,1)$ scalars and depend on the
gauge coupling constants and the $SU(1,1)$ angles,
$V_{ij}$ and $W_{ij}$ contain the
structure constants, depend on the matter
fields, and are symmetric resp.\ antisymmetric
in the indices $i,j$.

The extremum of the potential in the $\phi$ direction has been determined in
\cite{dRWP1}. For completeness we briefly review this
analysis in Appendix \ref{SU11}. The conclusion is that in the extremum
in the $SU(1,1)$ scalars the potential takes on the form
\begin{equation}
\label{pot1}
  V_0 = \sgn{C_-}\,\sqrt{\Delta}  - T_- \,,
\end{equation}
where (see \cite{dRWP1})
\begin{eqnarray}
\label{Cmin}
  C_{-} &=& \sum_{ij} g_ig_j\cos(\alpha_i-\alpha_j)V_{ij}\,,\\
\label{Tmin}
  T_{-}   &=& \sum_{ij} a_{ij} W_{ij} \,,\\
\label{Delta}
  \Delta &=& 2\,\sum_{ij}\sum_{kl} V_{ij}V_{kl} a_{ik}a_{jl}\,,\\
\label{Rij}
  R^{(ij)} &=& {\sgn C_-\over \sqrt{\Delta}}
      \sum_{kl} V_{kl}(2a_{ik}a_{jl} - a_{ij}a_{kl})\,,\\
\label{Iij}
  I^{(ij)} &=& -a_{ij}\,.
\end{eqnarray}
and
\begin{equation}
\label{aij}
   a_{ij}\equiv g_ig_j\sin(\alpha_i-\alpha_j)\,.
\end{equation}
The condition for this extremum is that $\Delta>0$, which implies that at least
two of the $SU(1,1)$ angles must be different. This in turn implies that
the gauge group must contain at least two simple subgroups.

\begin{center}
\renewcommand{\arraystretch}{1.5}
\begin{tabular}{|l||c|c|c|c||l|c|c|c|c|}
\hline
Group\hfil           & $-$ & $+$ & c & nc & Group\hfil & $-$ & $+$ & c & nc \\
\hline
$SO(3)_-$            & 3 & 0 & 3 & 0 &  $SO(3)_+$        & 0 & 3 & 3 & 0   \\
$SO(2,1)_-$          & 1 & 2 & 1 & 2 &  $SO(2,1)_+$      & 2 & 1 & 1 & 2   \\
$SO(3,1)_-$  & $3_c$ & $3_{nc}$ & 3 & 3 & $SO(3,1)_+$    & $3_{nc}$ & $3_c$ & 3 & 3 \\
$SL(3,\mathbb{R})_-$& 3 & 5 & 3 & 5 &
$SL(3,\mathbb{R})_+$& 5 & 3 & 3 & 5 \\
$SU(2,1)_-$& $4_c$ & $4_{nc}$ & 4 & 4 &
  $SU(2,1)_+$& $4_{nc}$ & $4_c$ & 4 & 4 \\
\hline
\end{tabular}
\renewcommand{\arraystretch}{1.0}
\end{center}
\vspace{.25truecm}
\noindent {\bf Table\ 1.} \ \
List of allowed simple groups. The first two columns
indicate how the group is embedded in $SO(6,6)$ with respect
to the signs in the metric $\eta_{RS}$, the column $c$ and $nc$
indicate the number of compact and noncompact generators,
respectively. The structure constants
of these groups are presented in Appendix \ref{SCS}.\\

Let us now make a list of possible simple subgroups of $G$, and discuss their
embedding in $SO(6,6)$.
The metric $\eta_{RS}$ is
the invariant metric of the global symmetry group $SO(6,6)$, which acts
on the fields in the fundamental representation. The gauge group $G$ acts
in the adjoint representation, and has 12 or less generators
(for $n=6$).
So the adjoint representation of $G$ must fit into the
fundamental representation of $SO(6,6)$. The generators $T_R$
of the simple subgroups of $G$ in the
fundamental representation are chosen
in such a way that $g_{RS}\equiv\tr T_RT_S=
\pm 2\delta_{RS}$,
with positive entries for the compact, and negative entries
for the noncompact directions.
The embedding of $G$ in $SO(6,6)$ is such that the metric
$g_{RS}$ coincides, up to an overall factor $\pm 2$,
with $\eta_{RS}$.
The factor 2 we absorb in the coupling constant
of the corresponding gauge group (see Appendix \ref{SCS} for
further properties of these groups).
With these restrictions one can easily
list all allowed simple factors $G_i$
of $G$. These are presented
in Table 1. Note that groups of dimension 10 and higher
are excluded since they do not
leave enough room for a second nonabelian subgroup.

The starting point for the remaining analysis is then (\ref{pot}). The
ingredients are
\begin{eqnarray}
  \label{Vij}
  V_{ij} &=& \bigg(-\noverm{1}{12}Z^{RU}Z^{SV}Z^{TW}
    + \quart Z^{RU}Z^{SV} (\eta^{TW} + Z^{TW})\bigg)\,
          f^{(i)}{}_{RST} f^{(j)}{}_{UVW}\,,
\\
\label{Wij}
  W_{ij} &=& \noverm{1}{36}\epsilon^{abcdef}
      Z_a{}^RZ_b{}^S Z_c{}^T Z_d{}^U Z_e{}^V Z_f{}^W\,
      f^{(i)}{}_{RST} f^{(j)}{}_{UVW}\,,
\end{eqnarray}
where $Z^{RS}=Z_a{}^RZ_a{}^S$. It is important to remember that the $Z_a{}^R$
are not the independent scalars, due to the constraint (\ref{conZ}). A useful
parametrisation is given in terms of $6\times 6$ matrices $G$ (symmetric) and
$B$ (antisymmetric).  We split the
indices $R,S,\ldots$ of $\eta_{RS}$ in $A,B,\ldots=1,\ldots,6$,
($\eta_{AB}=-\delta_{AB}$) and $I,J,\ldots=7,\ldots,12$,
($\eta_{IJ}=+\delta_{IJ}$). The scalar constraint (\ref{conZ})
then reads
\begin{equation}
   X X^T - Y Y^T = \unitm{6}\,,
\end{equation}
where $X_{a}{}^A = Z_a{}^A,\ Y_a{}^{I-6} = Z_a{}^I$, which is solved by
\begin{eqnarray}
\label{Xdef}
  X &=& {1\over 2}\left(
           G + G^{-1} + BG^{-1} - G^{-1}B - B G^{-1} B
      \right)\,, \\
\label{Ydef}
  Y &=& {1\over 2}\left(
           G - G^{-1} - BG^{-1} - G^{-1}B - B G^{-1} B
       \right)\,.
\end{eqnarray}
In \cite{dRWP1} we limited ourselves to the case where
\begin{equation}
\label{matterpar}
  G = \left(\begin{matrix}
            a \unitm{3} & 0 \\
            0  & a \unitm{3}
        \end{matrix}\right)\quad (a>0)\,,\qquad
  B = \left(\begin{matrix}
            0 & b \unitm{3} \\
        -b \unitm{3} & 0
        \end{matrix}\right) \,,
\end{equation}
in this paper we will use the complete $G$ and $B$.

We will analyse the potential at the point
$Z_0$, the origin of the scalar manifold,
given by $G=\unitmatrixDT,\ B=0$.
We will see in Section
\ref{firstder} that for many gauge groups the origin corresponds to an extremum
of the potential in all directions. At $Z=Z_0$ we have
\begin{eqnarray}
   Z_{0\,a}{}^R &=&
    \left(\begin{matrix} \unitm{6} & 0 \end{matrix}\right)\,,
   \quad
   Z_0{}^{RS}  =
    \left(\begin{matrix} \unitm{6} & 0 \cr
              0    & 0 \end{matrix}\right)\,,
   \quad   (\eta+Z)_0{}^{RS} =
 \left(\begin{matrix}0 & 0 \cr
                          0 & \unitm{6} \end{matrix}\right)\,.
\end{eqnarray}

Consider $V_{ij}$, as given in (\ref{Vij}).
The first term contains a product of three $Z_0{}^{RS}$, which are
diagonal and only
non-vanishing if all of $RSTUVW$ are in the range $1\ldots 6$.
The second term containing $\eta+Z$
is also diagonal with the two indices in the range $7\ldots 12$.
Therefore in $V_{ij}$ the indices of the structure constants of
$G_i$ and $G_j$ are contracted, implying that
they belong to the same factor in the product of groups. Therefore
$V_{ij}=0$ for $i\ne j$.

$W_{ij}$ can only be nonzero if there are subgroups $G_i$ and $G_j$ of $G$,
$i\ne j$, such that both have three generators in the range $1\ldots 6$
with structure constants $f_{ABC}$. These $G_i$ must therefore be
$SO(3)$ or $SU(2)$ subgroups of $G_i$ and $G_j$.

Therefore we have:
\begin{eqnarray}
   V_{0\,ii}   &=& -\noverm{1}{12} \,f^{(i)}{}_{ABC}f^{(i)}{}_{ABC}
                   +\noverm{1}{4}  \,f^{(i)}{}_{ABI}f^{(i)}{}_{ABI}\,,
                        \nonumber\\
               &=& -\half\qquad {\rm for}\ SO(3)_-,\ SO(3,1)_-,\ SL(3,\mathbb{R})_-\,,
               \nonumber\\
               && -2 \qquad {\rm for}\ SU(2,1)_-\,,
               \nonumber\\
               &&  \half\qquad {\rm for}\ SO(2,1)_+ \,,
              \nonumber\\
\label{V0ij}
               &&   \tfrac{3}{2}\qquad   {\rm for}\ SO(3,1)_+\,,
               \\
               &&   \tfrac{15}{2}\qquad  {\rm for}\ SL(3,\mathbb{R})_+\,,
               \nonumber\\
               &&   6\qquad   {\rm for}\ SU(2,1)_+\,,
               \nonumber\\
               &&   0\qquad   {\rm for}\ SO(2,1)_-,\ SO(3)_+\,,
               \nonumber\\
   V_{0\,ij} &=& 0 \qquad i\ne j\,,
               \nonumber\\
   W_{0\,ij} &=& 1 \qquad {\rm for\ subgroups}\ SO(3)_-,\ SO(3,1)_-,\
               SL(3,\mathbb{R})_-\,,
               \nonumber\\
               &=& 0 \qquad {\rm otherwise}\,.\nonumber
\end{eqnarray}
Since $V_{0\,ij}=0$ for $i\ne j$ we can simplify
(\ref{pot1}). We find
\begin{eqnarray}
\label{C0min}
  C_{0-} &=& \sum_{i} g_i^2 V_{0\,ii}\,,\\
\label{T0min}
  T_{0-} &=& \sum_{ij} a_{ij}W_{0\,ij} \,,\\
\label{Del0}
 \Delta_0 &=& 2\,\sum_{ij} \,
   V_{0\,ii}V_{0\,jj} a_{ij}^2\,,\\
\label{R0ii}
 R^{(0\,ii)} &=& {2\,\sgn C_{0-}\over \sqrt{\Delta_0}}
   \sum_j V_{0\,jj} a_{ij}^2 \,.
\end{eqnarray}
The off-diagonal $R^{ij}$ can also be nonzero, but they do not
appear in the potential (or its first and second derivatives, as we shall see)
in $Z_0$ because $V_{0\,ij}$ is diagonal.

Further restrictions come from the value of $V_{0\,ii}$. To have $\Delta>0$,
we must have at least two subgroups $G_i$ for which $V_{ii}\ne 0$. Note for
instance that the groups $SO(3)_+$ and $SO(2,1)_-$ have zero $V_{0\,ii}$, and
do not contribute to $\Delta$ in $Z_0$.
More restrictions will come from the requirement that
$Z_0$ corresponds to an extremum of the
potential.

There are two ways to make $V_0$ positive. One is to have $C_->0$, to which
 groups with positive $V_{0\,ii}$,
such as $SO(2,1)_+$ or $SO(3,1)_+$, contribute. A further positive
contribution can come from $T_-$, if $W_{ij}$ is nonzero and the $SU(1,1)$ angles
are appropriately chosen. This can be done for groups with $SO(3)_-^2$ as
a subgroup.

\subsection{The first derivatives of the potential\label{firstder}}

The unconstrained, independent scalar fields in the 12 vector multiplets are
21 components of the symmetric matrix $G$, and 15 components of the antisymmetric
matrix $B$. It is convenient to introduce a single matrix $P=(G+B)$ for
the 36 independent scalars. In Appendix \ref{PGB} we give the
derivatives of $X$ and $Y$ in $Z_0$ with respect to these parameters.
The first derivatives of $V$ and $W$ with respect to $P$ are:
\begin{eqnarray}
  {\partial V_{ij}\over \partial P} &=&
     \half\big( {\partial Z_f{}^R\over \partial P}Z_f{}^U +
                     Z_f{}^R {\partial Z_f{}^U\over \partial P} \big)
                 \, Z^{SV}(\eta+Z)^{TW}  f^{(i)}{}_{RST} f^{(j)}{}_{UVW}\,,
         \\
  {\partial W_{ij}\over \partial P} &=&
     \noverm{1}{12}\epsilon^{abcdef}\big(
           {\partial Z_a{}^R\over \partial P}Z_b{}^SZ_c{}^TZ_d{}^U Z_e{}^V Z_f{}^W
          +
      \nonumber\\
 &&\qquad + Z_a{}^RZ_b{}^SZ_c{}^T{\partial Z_d{}^U\over \partial P} Z_e{}^V Z_f{}^W
          \big)\,f^{(i)}{}_{RST} f^{(j)}{}_{UVW} \,.
\end{eqnarray}
Now we evaluate this in $Z_0$, using the fact that the derivative of $X$
with respect to $P$ vanishes in $Z_0$. The derivatives of $V$ and $W$
in $Z_0$ are then:
\begin{eqnarray}
\label{dV0}
  {\partial V_{ii}\over \partial P}\bigg|_0 &=&
     {\partial Z_A{}^I\over \partial P}\bigg|_0
             f^{(i)}{}_{IBJ} f^{(i)}{}_{ABJ}\,,\quad
   {\partial V_{ij}\over \partial P}\bigg|_0 =0\ {\rm for}\ i\ne j\,,
         \\
\label{dW0}
  {\partial W_{ij}\over \partial P}\bigg|_0 &=&
     \noverm{1}{12}\epsilon^{ABCDEF}
           {\partial Z_A{}^I\over \partial P}\bigg|_0
           \,f^{(i)}{}_{IBC} f^{(j)}{}_{DEF} - (i\leftrightarrow j)\,.
\end{eqnarray}
In the derivatives of $V_{ij}$
there are always contractions between the different groups, implying
that the derivatives vanish for $i\ne j$.
For the derivative of $V_{ii}$ to be nonzero
we see that $G_i$ must have two indices
$AB$ and two indices $IJ$.
This is not the case for the groups in Table 1.

The derivative of $W_{ij}$ is nonzero only if one of the groups $G_j$
or $G_i$
has an $SO(3)$ subgroup on the indices $DEF$, and the other has an
$SO(2,1)$ subgroup for which the $IBC$ structure
constants are nonzero. Groups $G$ with an $SO(3)_-\otimes SO(2,1)_+$
subgroup are therefore excluded.

If we require that $Z_0$ is an extremum of the potential, and
that $\Delta>0$ in the extremum is possible for a suitable choice
of the parameters, the number of allowed groups becomes sufficiently
small to make a complete analysis possible. The list of allowed groups
is given in Table~2.

To illustrate  Table~2, let's consider the group $SO(3,1)$.
The commutation relations for the $SO(3,1)$ Lie algebra can be
found in Appendix \ref{SCS}. The generators of the rotation subgroup are
denoted by $T$, the boosts by $K$.
To embed the adjoint of $SO(3,1)$ in $SO(6,6)$ we have two choices:
either the three $T$ correspond to the negative, and
the boosts $K$ to the positive entries of
$\eta_{RS}$ ($SO(3,1)_-$), or the other way round
($SO(3,1)_+$). If we choose $SO(3)_-\otimes SO(3,1)_-$,
the nonzero contributions to $V_{0\,ii}$ (\ref{V0ij}) come from $SO(3)_-$
and from the rotation subgroup of $SO(3,1)_-$, both contribute $-\half$,
and $\Delta_0$ (\ref{Del0}) is positive. The structure constants
$f_{ABI}$ vanish, and do not contribute to $V_0$
in this case. For the same reason, the first derivatives
of $V$ and $W$ (\ref{dV0},\ref{dW0}) vanish. So this case is interesting,
and will appear as a subgroup of the groups considered in
Section \ref{G6} and \ref{G7}.
On the other hand, if we choose $SO(3)_-\otimes SO(3,1)_+$ then the boosts
are in the $ABC$ range of the indices, and structure constants $f_{ABI}$
are nonzero. The potential $V_{0\,ii}$ gets contributions from
$SO(3)_-$, and now the second term in (\ref{V0ij}) will contribute positively.
This will make $\Delta_0$ negative. Also the first derivative of $W$ will
be nonzero, so this group cannot be used for our purposes.

\begin{center}
\renewcommand{\arraystretch}{1.5}
\begin{tabular}{||l|l||}
\hline
Groups  & Properties                   \\
\hline
$SO(3)_-^2\otimes SO(3)_+^2$   &\hfil  \\
$SO(3)_-\otimes SO(2,1)_+\otimes SO(2,1)_-\otimes SO(3)_+$
          &$\Delta_0<0$, no extremum\\
$SO(3)_-\otimes SO(2,1)_-^3$ & $\Delta_0=0$ \\
$SO(2,1)_+^3\otimes SO(3)_+$  & \hfil \\
$SO(2,1)_+^2  \otimes SO(2,1)_-^2$ & \hfil \\
$SO(3,1)_+\otimes SO(2,1)_+\otimes SO(2,1)_-$ & \hfil\\
$SO(3,1)_+\otimes SO(3,1)_+$ & \hfil \\
$SO(3,1)_+ \otimes SO(3)_- \otimes  SO(3)_+$
          & $\Delta_0<0$, no extremum       \\
$SO(3,1)_-\otimes SO(2,1)_+\otimes SO(2,1)_-$ & $\Delta_0<0$, no extremum\\
$SO(3,1)_-\otimes SO(3,1)_-$ & \hfil \\
$SO(3,1)_-\otimes SO(3,1)_+$ & $\Delta_0<0$, no extremum \\
$SO(3,1)_- \otimes SO(3)_-\otimes SO(3)_+$
          &\hfil       \\
$SL(3,{\mathbb{R}})_+\otimes SO(3)_+$ & $\Delta_0=0$ \\
$SL(3,{\mathbb{R}})_-\otimes SO(3)_-$ & \hfil \\
$SL(3,{\mathbb{R}})_+\otimes SO(2,1)_-$ & $\Delta_0=0$ \\
$SL(3,{\mathbb{R}})_-\otimes SO(2,1)_+$ & $\Delta_0<0$, no extremum \\
$SU(2,1)_+\otimes SO(2,1)_+$ &  \hfil \\
$SU(2,1)_+\otimes SO(2,1)_-$ &  $\Delta_0=0$ \\
$SU(2,1)_-\otimes SO(2,1)_+$ &  $\Delta_0<0$, no extremum \\
$SU(2,1)_-\otimes SO(2,1)_-$ &  $\Delta_0=0$ \\
\hline
\end{tabular}
\renewcommand{\arraystretch}{1.0}
\end{center}
\vspace{.25truecm}
\noindent {\bf Table\ 2.} \ \
List of possible gauge groups $G$.
Nine groups have an extremum with respect to
the matter scalars in $Z_0$ with positive $\Delta_0$. \\

\subsection{The second derivatives of the potential\label{secondder}}

The full potential is given in (\ref{pot}),
and its second derivatives are:
\begin{eqnarray}
\label{DDphi}
 {\partial^2 V\over \partial\phi^2} &=&
   \sum_{ij} {\partial^2 R^{(ij)}\over \partial\phi^2} V_{ij}\,,
 \\
 \label{DDphiZ}
  {\partial^2 V\over \partial\phi\partial P} &=&
   \sum_{ij} {\partial R^{(ij)}\over \partial\phi}
             {\partial V_{ij}\over \partial P} \,,
 \\
\label{DDZZ}
  {\partial^2 V\over \partial P^2} &=&
   \sum_{ij} R^{(ij)} {\partial^2 V_{ij}\over \partial P^2}
             +I^{(ij)}{\partial^2 W_{ij}\over \partial P^2}\,.
\end{eqnarray}
The second derivatives (\ref{DDphi}) were studied in \cite{dRWP1}. The sign of
(\ref{DDphi}) depends on the sign of $C_-$. For positive (negative) $C_-$ the
extremum in the $SU(1,1)$ scalars is a minimum (maximum).
The mixed second derivatives vanish if either the derivatives with respect to
the $SU(1,1)$ scalars $\phi$ or with respect to the matter scalars vanishes.
In this section we will evaluate the second derivatives (\ref{DDZZ})
with respect to the matter scalars in $Z_0$.

We therefore calculate the second derivatives of $V_{ij}$ and $W_{ij}$ with
respect to the independent scalars $P$.
They are:
\begin{eqnarray}
  {\partial^2 V_{ij}\over \partial P_{1}\partial P_{2}} &=&
  \half\bigg\{
   \bigg({\partial^2 Z_f{}^R\over \partial P_{1}\partial P_{2}}Z_f{}^U +
            Z_f{}^R {\partial^2 Z_f{}^U\over \partial P_{1}\partial P_{2}}
   \bigg)\,Z^{SV}(\eta+Z)^{TW} +\nonumber\\
  && + \bigg({\partial Z_f{}^R\over \partial P_{1}}
              {\partial Z_f{}^U\over \partial P_{2}}+
              {\partial Z_f{}^R\over \partial P_{2}}
              {\partial Z_f{}^U\over \partial P_{1}}
   \bigg)\,Z^{SV}(\eta+Z)^{TW} +\nonumber\\
  && + \bigg({\partial Z_f{}^R\over \partial P_{1}}Z_f{}^U +
            Z_f{}^R {\partial Z_f{}^U\over \partial P_{1}}\bigg)\,
       \bigg({\partial Z_g{}^S\over \partial P_{2}}Z_g{}^V +
            Z_g{}^S {\partial Z_g{}^V\over \partial P_{2}}\bigg)
       \,(\eta+Z)^{TW} +\nonumber\\
  && + \bigg({\partial Z_f{}^R\over \partial P_{1}}Z_f{}^U +
            Z_f{}^R {\partial Z_f{}^U\over \partial P_{1}}\bigg)\,
           Z^{SV}\,
       \bigg({\partial Z_g{}^T\over \partial P_{2}}Z_g{}^W +
            Z_g{}^T {\partial Z_g{}^W\over \partial P_{2}}\bigg)
      \bigg\}\times
   \nonumber\\
\label{DDVDT}
  && \times f^{(i)}{}_{RST} f^{(j)}{}_{UVW} \,,\\
   {\partial^2 W_{ij}\over \partial P_{1}\partial P_{2}} &=&
   \noverm{1}{12}\,\epsilon^{a_1\ldots a_6} \bigg\{
      {\partial^2 Z_{a_1}{}^R\over \partial P_{1}\partial P_{2}}
      Z_{a_2}{}^SZ_{a_3}{}^T Z_{a_4}{}^U Z_{a_5}{}^V Z_{a_6}{}^W +
   \nonumber\\
   &&
    + 2{\partial Z_{a_1}{}^R\over \partial P_{1}}
      {\partial Z_{a_2}{}^S\over \partial P_{2}}
       Z_{a_3}{}^T Z_{a_4}{}^U Z_{a_5}{}^V Z_{a_6}{}^W +
   \nonumber\\
   &&
      + 3 {\partial Z_{a_1}{}^R\over \partial P_{1}}
       Z_{a_2}{}^SZ_{a_3}{}^T
       {\partial Z_{a_4}{}^U\over \partial P_{2}}Z_{a_5}{}^V Z_{a_6}{}^W
 \nonumber\\
 \label{DDWND}
 &&
  - (RST \leftrightarrow UVW) \bigg\}
       f^{(i)}{}_{RST} f^{(j)}{}_{UVW} \,.
\end{eqnarray}
We now evaluate both expressions for $Z=Z_0$. From
(\ref{DDVDT}) it is clear that in $Z=Z_0$ we will get
contractions between the two structure constants, so that they must belong to
the same subgroup. Therefore only the second derivatives of $V_{ii}$ are nonzero
in $Z_0$.
We find:
\begin{eqnarray}
 && {\partial^2 V_{ii}\over \partial P_{1}\partial P_{2}} \bigg|_0 =
  \bigg\{
    \, {\partial^2 Z_A{}^R\over \partial P_{1}\partial P_{2}}\bigg|_0
                  \, f^{(i)}{}_{RBJ} f^{(i)}{}_{ABJ}
      +  {\partial Z_f{}^I\over \partial P_{1}}\bigg|_0
              {\partial Z_f{}^J\over \partial P_{2}}\bigg|_0
           \,  f^{(i)}{}_{IBK} f^{(i)}{}_{JBK} +
           \nonumber\\
  &&\quad + \bigg({\partial Z_A{}^I\over \partial P_{1}}\bigg|_0
             {\partial Z_B{}^J\over \partial P_{2}}\bigg|_0
               f^{(i)}{}_{IJK} f^{(i)}{}_{ABK}
      + {\partial Z_A{}^I\over \partial P_{1}}\bigg|_0
           {\partial Z_B{}^J\over \partial P_{2}}\bigg|_0
               f^{(i)}{}_{AJK} f^{(i)}{}_{IBK}  \bigg)
           \nonumber\\
  \label{DDVDT0}
  &&\quad + \bigg({\partial Z_A{}^I\over \partial P_{1}}\bigg|_0
             {\partial Z_B{}^J\over \partial P_{2}}\bigg|_0
               f^{(i)}{}_{ICJ} f^{(i)}{}_{ACB}
                   +
           {\partial Z_A{}^I\over \partial P_{1}}\bigg|_0
             {\partial Z_B{}^J\over \partial P_{2}}\bigg|_0
               f^{(i)}{}_{ICB} f^{(i)}{}_{ACJ}  \bigg\} \,,
          \nonumber\\
 &&{\partial^2 W_{ij}\over \partial P_{1}\partial P_{2}}\bigg|_0 =
   \noverm{1}{12}\epsilon^{ABCDEF} \bigg(
      {\partial^2 Z_{A}{}^R\over \partial P_{1}\partial P_{2}}\bigg|_0
             f^{(i)}{}_{RBC} f^{(j)}{}_{DEF}
   \nonumber\\
   &&\quad +\, 2\,
   {\partial Z_{A}{}^I\over \partial P_{1}}\bigg|_0
      {\partial Z_{B}{}^J\over \partial P_{2}}\bigg|_0
       f^{(i)}{}_{IJC} f^{(j)}{}_{DEF}
   \nonumber\\
   \label{DDWND0}
   &&\quad
     + \,3\,
      {\partial Z_{A}{}^I\over \partial P_{1}}\bigg|_0
          {\partial Z_{D}{}^J\over \partial P_{2}}\bigg|_0
       f^{(i)}{}_{IBC} f^{(j)}{}_{JEF}
  - (i \leftrightarrow j)  \bigg)  \,.
\end{eqnarray}
The second derivatives of $Z$ ($X$ and $Y$) are given in Appendix \ref{PGB}.

\subsection{Masses of scalar and vector fields\label{scalarmass}}

The mass-matrix should be normalised in relation to the kinetic terms.
The kinetic term of the matter scalar fields $Z$ is independent of the
gauging - the gauge fields occur only in the covariantisations. The
kinetic term of the $Z_a{}^R$ reads (ignoring the gauge field contributions)
\begin{equation}
   -\half\eta_{RS}\partial_\mu Z_a{}^R\partial^\mu Z_a{}^S
     -\noverm{1}{8}\eta_{RS}\eta_{TU}
    Z_a{}^R \overleftrightarrow{\partial_\mu}Z_b{}^S\,
    Z_a{}^T \overleftrightarrow{\partial^\mu}Z_b{}^U \,.
\end{equation}
This should be evaluated in $Z=Z_0$ and expressed in terms of
$P_{ab}$, giving
\begin{equation}
   -\half \partial_\mu P_{ab}\partial^\mu P_{ab}
   \,,
\end{equation}
the standard normalization for scalar fields. Together with the contribution
from the potential we therefore get
\begin{equation}
    -\half \partial_\mu P_{ab}\partial^\mu P_{ab}
    - V_0
     -\half (P_{ab}-\delta_{ab}) (P_{cd}-\delta_{cd})
     {\partial^2 V\over \partial P_{ab} \partial P_{cd}}\bigg|_0 \,.
\end{equation}
On shifting the scalar fields in the kinetic term by the constant $\delta_{ab}$
we see that the matrix of second derivatives
we have calculated in the previous section is precisely the mass-matrix.

The mass-matrix for the $SU(1,1)$ scalars was given is given in \ref{actionphi}
in Appendix \ref{SU11}.
The kinetic and mass terms for these scalars are:
\begin{equation}
    -\half (\partial_\mu x'\partial^\mu x' + \partial_\mu y'\partial^\mu y')
    - \half\sgn C_-\,\sqrt{\Delta_0}(x^{\prime\,2}+ y^{\prime\,2})\,.
\end{equation}
In the case $C_-<0$ we have two tachyons. The relation between $\phi_{1,2}$ and
$x$ and $y$ is explained in Appendix \ref{SU11}.

The vector masses follow from the coupling of the vectors to the scalars
$Z$. We have in the covariant derivative:
\begin{equation}
 {\cal D}_\mu Z_a{}^{R} =
  \partial_\mu Z_a{}^R - V_{\mu\,ab} Z_b{}^R
    - A_\mu{}^S g f_{ST}{}^R Z_a{}^T\,.
\end{equation}
After elimination of $V$ the scalar kinetic term becomes:
\begin{equation}
   -\half\eta_{RS} D_\mu Z_a{}^R D^\mu Z_a{}^S
     -\noverm{1}{2}\eta_{RS}\eta_{TU}
    Z_a{}^R Z_a{}^T {D_\mu}Z_b{}^S {D^\mu}Z_b{}^U \,,
\end{equation}
where $D$ contains the gauge field $A$ only. Now substitute $Z=Z_0$
and isolate the $A^2$ terms. The result is, after writing out the
indices $R,S$ in terms of $A,B$ and $I,J$:
\begin{equation}
   -\half g_i^2 A_\mu{}^AA^{\mu\,B} f^{(i)}{}_{ACI}f^{(i)}{}_{BCI}
      - g_i^2 A_\mu{}^AA^{\mu\,K}   f^{(i)}{}_{ACI}f^{(i)}{}_{KCI}
   -\half g_i^2 A_\mu{}^KA^{\mu\,L} f^{(i)}{}_{KCI}f^{(i)}{}_{LCI}\,.
\end{equation}
We see that there are vector masses only for noncompact groups.
That is to be expected, since these are the noncompact generators
do not leave $Z_0$ invariant. The second term vanishes for all groups
considered.
The first term get contributions from gauge groups with an $SO(2,1)_+$
subgroups,
the last one from $SO(2,1)_-$ subgroups. The masses are
proportional to the corresponding $g_i^2$, and independent of the
$SU(1,1)$ angles.

\section{Semi-simple gauge groups\label{groups}}

Table 2 left us with nine allowed groups. In this Section we will discuss
these nine cases separately. In Table 3 we give a list of the allowed groups,
with their basic properties. The sign of $C_-$ is important, because
it determines the behaviour of the $SU(1,1)$ scalars: if $C_-<0$, the $SU(1,1)$
scalars are at a maximum, if $C_->0$ at a minimum. Clearly a stable
minimum in all 38 scalar directions requires $C_->0$.
We now discuss the nine groups in the order of Table~3.

\begin{center}
\renewcommand{\arraystretch}{1.5}
\begin{tabular}{||l|l|l||}
\hline
Groups  &  $C_-$   &  $V_0$   \\
\hline
$SO(2,1)_+^3\otimes SO(3)_+$  & $\half{(g_1^2+g_2^2+g_3^2)}$ &
       $\sqrt{a_{12}^2+a_{13}^2+a_{23}^2}$\\
$SO(2,1)_+^2  \otimes SO(2,1)_-^2$ & $\half(g_1^2+g_2^2)$ & $|a_{12}|$\\
$SO(3,1)_+\otimes SO(2,1)_+\otimes SO(2,1)_-$ & $\half(3g_1^2+ g_2^2)$ &
       $\sqrt{3}|a_{12}|$\\
$SO(3,1)_+\otimes SO(3,1)_+$ & $\tfrac{3}{2}(g_1^2+g_2^2)$ & $3|a_{12}|$\\
$SO(3)_-^2\otimes SO(3)_+^2$   & $-\half (g_1^2+g_1^2)$ &
        $-|a_{12}|-2a_{12}$\\
$SO(3)_-\otimes SO(3,1)_- \otimes SO(3)_+$ &  $-\half (g_1^2+g_1^2)$&
        $-|a_{12}|-2a_{12}$     \\
$SO(3,1)_-\otimes SO(3,1)_-$ & $-\half (g_1^2+g_2^2)$ & $-|a_{12}| -2a_{12}$\\
$SL(3,\mathbb{R})_-\otimes SO(3)_-$ & $-\half (g_1^2+g_2^2)$ & $-|a_{12}| -2a_{12}$\\
$SU(2,1)_+\otimes SO(2,1)_+$ & $ 6g_1^2+\half g_2^2$ & $2\sqrt{3}|a_{12}|$\\
\hline
\end{tabular}
\renewcommand{\arraystretch}{1.0}
\end{center}
\vspace{.25truecm}
\noindent {\bf Table\ 3.} \ \
List of possible gauge groups $G$. $V_0$ is the value of the potential in
$Z=Z_0$. $C_->0$ (${}<0$) implies that the $SU(1,1)$ scalars are at a minimum
(maximum). \\

In the following Sections \ref{G1}-\ref{G9}
we will discuss the nine gauge groups which
have $\Delta_0>0$ and an extremum for the matter scalars in the origin
of the scalar manifold. We present for each case the mass-matrix for
the matter scalars. The masses of the $SU(1,1)$ scalars are always
given by $\sgn C_-\sqrt{\Delta_0}$, indicating that they are always both
tachyonic or both positive.

For the first case, the group $SO(2,1)_+^3\otimes SO(3)_+$ presented
in Section \ref{G1}, we will give full details of the analysis.
For the other cases the procedure should then be clear, and we will
limit ourselves to the presentation of the results.

\subsection{$SO(2,1)_+^3\otimes SO(3)_+$\label{G1}}

In this case the groups have to be arranged as follows:
\begin{equation}
R,S,\ldots = \overbrace{\,1\ 2\ 7\,}^{i=1}\ \overbrace{\,3\ 4\ 8\,}^{i=2}\
\overbrace{\,5\ 6\ 9\,}^{i=3}\ \overbrace{\,10\ 11\ 12\,}^{i=4}\,.
\end{equation}
Then all $V_{0\,ii}=\half$ for $i=1,2,3$, while $V_{044}=0$,
and we have:
\begin{eqnarray}
\label{C0min1}
  C_{0-} &=& \half(g_1^2+g_2^2+g_3^2)\,,
    \\
 \label{T0min1}
  T_{0-}  &=& 0\,,
    \\
\label{Del01}
  \Delta_0 &=& a_{12}^2+a_{13}^2+a_{23}^2\,,
    \\
\label{R0ii1}
  R^{(0\,ii)} &=&
   {1 \over\sqrt{\Delta_0}}\,\big(
    a_{i1}^2 + a_{i2}^2 + a_{i3}^2 \big)\,.
\end{eqnarray}
giving
\begin{eqnarray}
\label{V01}
  V_0 &=& \sqrt{\Delta_0}\,,\\
\label{dV01}
  {\partial V\over \partial P_{ab}}\bigg|_0 &=& 0 \,,\\
\label{ddV01}
  {\partial^2 V\over \partial P_{ab}\partial P_{cd}}\bigg|_0 &=&
    {1 \over\sqrt{\Delta_0}}\,\sum_{i=1}^3
    \big(a_{i1}^2 + a_{i2}^2 + a_{i3}^2 \big)
   {\partial^2 V_{ii}\over \partial P_{ab}\partial P_{cd}}\bigg|_0
  - \sum_{i<j} 2a_{ij}
  {\partial^2 W_{ij}\over \partial P_{ab}\partial P_{cd}}\bigg|_0\,.
\end{eqnarray}
Notice that the group $SO(3)_+$ does not contribute to the second
derivatives, structure constants with indices $IJK$ appear in
(\ref{DDVDT0}) in the third term, but in combination with
a structure constant $f_{ABK}$, which $SO(3)_+$ does not have.
The non-vanishing second derivative matrices required for calculating
the mass-matrix are:
\begin{eqnarray}
  {\partial^2 V_{11}\over \partial P_{ab}\partial P_{cd}} \bigg|_0 &=&
     (\delta_{ac}-\delta_{a1}\delta_{c1})
      (\delta_{b1}\delta_{d1}+\delta_{b2}\delta_{d2})\,,
      \nonumber\\
   {\partial^2 V_{22}\over \partial P_{ab}\partial P_{cd}} \bigg|_0 &=&
     (\delta_{ac}-\delta_{a2}\delta_{c2})
      (\delta_{b3}\delta_{d3}+\delta_{b4}\delta_{d4})\,,
      \nonumber\\
    {\partial^2 V_{33}\over \partial P_{ab}\partial P_{cd}} \bigg|_0 &=&
     (\delta_{ac}-\delta_{a3}\delta_{c3})
      (\delta_{b5}\delta_{d5}+\delta_{b6}\delta_{d6})\,,
      \nonumber\\
    {\partial^2 W_{12}\over \partial P_{ab}\partial P_{cd}} \bigg|_0 &=&
     (\delta_{a1}\delta_{c2}-\delta_{a2}\delta_{c1})
      (\delta_{b5}\delta_{d6}-\delta_{b6}\delta_{d5})\,,
      \nonumber\\
     {\partial^2 W_{13}\over \partial P_{ab}\partial P_{cd}} \bigg|_0 &=&
     (\delta_{a1}\delta_{c3}-\delta_{a3}\delta_{c1})
      (\delta_{b3}\delta_{d4}-\delta_{b4}\delta_{d3})\,,
      \nonumber\\
     {\partial^2 W_{23}\over \partial P_{ab}\partial P_{cd}} \bigg|_0 &=&
     (\delta_{a2}\delta_{c3}-\delta_{a3}\delta_{c2})
      (\delta_{b1}\delta_{d2}-\delta_{b2}\delta_{d1})\,.
\end{eqnarray}
The resulting eigenvalues of
${\partial^2 V\over \partial P_{ab}\partial P_{cd}}\bigg|_0$ are then
\begin{eqnarray}
   &&0\ (6\ \times),\ {1\over \sqrt{\Delta_0}}(a_{12}^2+a_{13}^2)\ (6\ \times),
                    \ {1\over \sqrt{\Delta_0}}(a_{13}^2+a_{23}^2)\ (6\ \times),
                    \ {1\over \sqrt{\Delta_0}}(a_{12}^2+a_{23}^2)\ (6\ \times)\,,
                    \nonumber\\
   &&{1\over \sqrt{\Delta_0}}(a_{12}^2+a_{13}^2)\pm 2a_{23}\ (2\ \times)\,,\
     {1\over \sqrt{\Delta_0}}(a_{13}^2+a_{23}^2)\pm 2a_{12}\ (2\ \times)\,,\
       \nonumber\\
   &&{1\over \sqrt{\Delta_0}}(a_{12}^2+a_{23}^2)\pm 2a_{13}\ (2\ \times)\,.
\end{eqnarray}
One can obtain these eigenvalues as follows.
The second derivatives of $V_{ii}$ are
diagonal in the parameters $P$, in the sense that always $a=c,\ b=d$. The list
of elements of $P$ giving a nonzero second derivative of $V_{ii}$ is
\footnote{We use the notation
$ab$ to indicate the second derivative with respect to $P_{ab}$
(diagonal elements of the $36\times 36$ second derivative matrix), and
$(ab,cd)$ for the second derivative with respect to $P_{ab}$ and $P_{cd}$
(off-diagonal elements of the second derivative matrix).}:
\begin{eqnarray}
&&V_{11}:\ {1\over \sqrt{\Delta_0}}(a_{12}^2+ a_{13}^2)\,
    [\,21,\ 22,\ 31,\ 32,\ 41,\ 42,\ 51,\ 52,\ 61,\ 62\,]\,,\nonumber\\
&&V_{22}:\ {1\over \sqrt{\Delta_0}}(a_{12}^2+ a_{23}^2)\,
    [\,13,\ 14,\ 33,\ 34,\ 43,\ 44,\ 53,\ 54,\ 63,\ 64\,]\,,\nonumber\\
&&V_{33}:\ {1\over \sqrt{\Delta_0}}(a_{13}^2+ a_{23}^2)\,
    [\,15,\ 16,\ 25,\ 26,\ 45,\ 46,\ 55,\ 56,\ 65,\ 66\,]\,.\nonumber
\end{eqnarray}
On the other hand, $W_{ij}$ has a non-vanishing second derivative for the
following pairs of elements of $P$:
\begin{eqnarray}
&&W_{12}:\ -2a_{12}
    [\,(15,\ 26),\ (16,\ 25),\ (25,\ 16),\ (26,\ 15)\,]\,,\nonumber\\
&&W_{13}:\ -2a_{13}
    [\,(13,\ 34),\ (14,\ 33),\ (33,\ 14),\ (34,\ 13)\,]\,,\nonumber\\
&&W_{23}:\ -2a_{23}
    [\,(21,\ 32),\ (22,\ 31),\ (31,\ 22),\ (32,\ 21)\,]\,.\nonumber
\end{eqnarray}
The six eigenvalues $(a_{13}^2+a_{23}^2)/\sqrt{\Delta_0}$
are associated with the six diagonal
elements of the second derivative matrix coming from $V_{11}$:
$41,\ 42,\ 51,\ 52,\ 61,\ 62$ are the corresponding elements
of $P$. The six eigenvalues $(a_{12}^2+a_{23}^2)/\sqrt{\Delta_0}$ and
$(a_{12}^2+a_{13}^2)/\sqrt{\Delta_0}$ arise in a similar way
from the derivatives of $V_{22}$ and $V_{33}$, respectively. The remaining
diagonal contributions from $V_{ii}$ combine with the corresponding
derivatives of $W_{ij}$: $V_{11}$ with $W_{23}$, etc. These elements
of $P$ give rise to $2\times 2$ submatrices in the
matrix of second derivatives, with the eigenvalues as indicated.
The zero eigenvalues of the second
derivative matrix are associated with the elements $11,\ 12,\ 23,\ 24,\
35,\ 36$ of $P$ which do not occur anywhere in the second derivatives.

The six zero eigenvalues indicate that the solution $Z=Z_0$ breaks
 the gauge symmetry. In each of the $SO(2,1)_+$ groups the
 $SO(2,1)$ symmetry is broken to $U(1)$. The six massless scalars give
 masses to the gauge vectors, as we have seen in Section \ref{scalarmass}.

The extremum we have obtained in this case is not stable.
Assume that all 36 eigenvalues are nonnegative.
If $(a_{12}^2+a_{13}^2)/\sqrt{\Delta_0}\pm 2a_{23}$
has to be positive then necessarily
\begin{equation}
   |a_{23}| < \sqrt{\Delta_0}(\sqrt{2}-1)\,.
\end{equation}
This has to be true then for $a_{13}$ and $a_{23}$ as well, implying
\begin{equation}
   \Delta_0  < 3\Delta_0 (\sqrt{2}-1)^2 < \Delta_0\,.
\end{equation}
Therefore our assumption that all eigenvalues are positive must be false.
In fact, there are two, four or six negative eigenvalues. To have six
negative eigenvalues choose $a_{12}=a_{23}=a_{13}=\sqrt{\Delta_0/3}$.

\subsection{$SO(2,1)_+^2\otimes SO(2,1)_-^2$\label{G2}}

The embedding of the subgroups is as follows:
\begin{equation}
R,S,\ldots = \overbrace{\,1\ 2\ 7\,}^{i=1}\ \overbrace{\,3\ 4\ 8\,}^{i=2}\
\overbrace{\,5\ 9\ 10\,}^{i=3}\ \overbrace{\,6\ 11\ 12\,}^{i=4}\,.
\end{equation}
Although the $V_{0\,ii}$  vanish for each $SO(2,1)_-$
factor, the second derivatives don't. We now have:
\begin{eqnarray}
\label{C0min2}
  C_{0-} &=& \half(g_1^2+g_2^2)\,,
  \nonumber\\
\label{T0min2}
  T_{0-} &=& 0\,,\\
\label{Del02}
  \Delta_0 &=& a_{12}^2
           \,, \\
\label{R0ii2}
  R^{(0\,ii)} &=&
   {1 \over |a_{12}|}\,(
      a_{i1}^2 + a_{i2}^2 )\,,\\
\label{V02}
  V_0 &=& |a_{12}|\,.
\end{eqnarray}

Now we find the following eigenvalues:
\begin{eqnarray}
   &&0\ (8\ \times),\ |a_{12}|\ (4\ \times),\ |a_{12}|+R^{(0\,33)}\ (8\ \times),\
     |a_{12}|+R^{(0\,44)}\ (8\ \times), \nonumber\\
   &&R^{(0\,33)}\ (2\ \times),\ R^{(0\,44)}\ (2\ \times),\ 2a_{12}\ (2\ \times),
     -2a_{12}\ (2 \times)\,.
\end{eqnarray}
$Z=Z_0$ breaks the gauge symmetry to $U(1)^4$, so the eight zero eigenvalues
correspond to the Goldstone bosons that produce the masses of the gauge fields.

There are two negative eigenvalues, proportional to $a_{12}$.
It does not help to set $\alpha_1=\alpha_2$ to eliminate them, since this
would make $\Delta=0$ and invalidate the analysis of
the $SU(1,1)$ scalars.

\subsection{$ SO(3,1)_+\otimes SO(2,1)_+\otimes SO(2,1)_-$\label{G3}}

In this case the groups are arranged as follows:
\begin{equation}
R,S,\ldots =
\overbrace{\,1\ 2\ 3\ 10\ 11\ 12\,}^{i=1}\
\overbrace{\,4\ 5\ 7\,}^{i=2}\
\overbrace{\,6\ 8\ 9\,}^{i=3}\,.
\end{equation}
The rotation subgroup of the $SO(3,1)$ subgroup is
embedded on the indices $10\ldots 12$, the boosts on the indices
$1\ldots 3$.
Here we have:
\begin{eqnarray}
\label{C0min3}
  C_{0-} &=& \half(3 g_1^2+ g_2^2)\,,
    \\
 \label{T0min3}
  T_{0-}  &=& 0\,,
    \\
\label{Del03}
  \Delta_0 &=& 3 a_{12}^2\,,
    \\
\label{R0ii3}
  R^{(0\,ii)} &=&
   {1 \over \sqrt{3}|a_{12}|}\,\big(
     3 a_{i1}^2 +  a_{i2}^2 \big)\,, \\
\label{V03}
  V_0 &=& \sqrt{3}|a_{12}|\,.
\end{eqnarray}

The complete list of eigenvalues in this case is:
\begin{eqnarray}
  &&0\ (7\ \times),\ 2|a_{12}|/\sqrt{3}\ (5\ \times),\
  \sqrt{3}|a_{12}|\ (6\ \times),\nonumber\\
  &&|a_{12}|(1+\sqrt{13})/\sqrt{3}\ (3\ \times),\
    |a_{12}|(1-\sqrt{13})/\sqrt{3}\ (3\ \times),\nonumber\\
  &&|a_{12}|(1+\sqrt{37})/\sqrt{3}\ (1\ \times),\
    |a_{12}|(1-\sqrt{37})/\sqrt{3}\ (1\ \times),\nonumber\\
  &&{1\over \sqrt{3}|a_{12}|}\,
     (3a_{12}^2 + 3a_{13}^2 + a_{23}^2)\ (4 \times),\
    {1\over \sqrt{3}|a_{12}|}\,
     (2a_{12}^2 + 3a_{13}^2 + a_{23}^2)\ (6 \times)\,.
\end{eqnarray}
Altogether then we find that this group gives rise to four negative eigenvalues
of the mass-matrix. The zero eigenvalues can again correspond to the
Goldstone bosons of the broken noncompact gauge symmetries. The negative eigenvalues
are proportional to $|a_{12}|$, which however we cannot set to zero because then
also $\Delta_0=0$.

\subsection{$SO(3,1)_+\otimes SO(3,1)_+$\label{G4}}

In this case the groups have to be arranged as follows:
\begin{equation}
R,S,\ldots =
\overbrace{\,1\ 2\ 3\ 10\ 11\ 12\,}^{i=1}\
\overbrace{\,4\ 5\ 6\ 7\ 8\ 9\,}^{i=2}\,.
\end{equation}
In this case the rotation subgroup of the two $SO(3,1)$ subgroups is
embedded on the indices $7\ldots 12$, the boosts on the indices
$1\ldots 6$.
Here $V_{0\,11}$ and $V_{0\,22}$ each are $\tfrac{3}{2}$, $W_{12}=0$.
and we have:
\begin{eqnarray}
\label{C0min4}
  C_{0-} &=& \tfrac{3}{2}\,(g_1^2+g_2^2)\,,
    \\
 \label{T0min4}
  T_{0-}  &=& 0\,,
    \\
\label{Del04}
  \Delta_0 &=& 9 a_{12}^2\,,
    \\
\label{R0ii4}
  R^{(0\,ii)} &=&
   {1 \over |a_{12}|}\,\big(
    a_{i1}^2 + a_{i2}^2 \big)\,, \\
  \label{V04}
  V_0 &=&  3|a_{12}|  \,.
\end{eqnarray}

The eigenvalues are:
\begin{eqnarray}
   0\ (15\ \times),\ 2|a_{12}|\ (10\ \times),\ 4|a_{12}|\ (9\ \times),\
   8|a_{12}|\ (1\ \times),\ -4|a_{12}|\ (1\ \times)\,.
\end{eqnarray}

In this case there is a single negative eigenvalue. There are
now more zero eigenvalues than the number required by
the breaking of gauge invariance.

\subsection{$SO(3)_-^2\otimes SO(3)_+^2$\label{G5}}

In the case of four $SO(3)$ groups these are arranged over the index values
$R,S = 1,\ldots 12$ as follows:
\begin{equation}
R,S,\ldots =
\overbrace{\,1\ 2\ 3\,}^{i=1}\
\overbrace{\,4\ 5\ 6\,}^{i=2}\
\overbrace{\,7\ 8\ 9\,}^{i=3}\
\overbrace{\,10\ 11\ 12\,}^{i=4} \,.
\end{equation}
This simplifies considerably the expressions derived in the Section \ref{Pot}
for the potential, and its first and second derivatives in the point $Z_0$.
The only terms which contribute are those for which the indices on the structure
constants are either $ABC$ (in the range $1\ldots 6$) or $IJK$ (in the range
$7\ldots 12$).
We find
\begin{eqnarray}
\label{C0min5}
  C_{0-} &=& -\half (g_1^2+g_2^2)\,,\\
\label{T0min5}
  T_{0-} &=& 2a_{12} \,,\\
\label{Delta05}
 \Delta_0 &=& a_{12}^2\,,\\
\label{R0ii5}
 R^{(0\,ii)} &=& {1\over |a_{12}|} \sum_i(a_{i1}^2+a_{i2}^2)\,,\\
 \label{V05}
   V_0 &=& -|a_{12}| - 2a_{12}\,.
\end{eqnarray}

The eigenvalues of the mass-matrix are
\begin{equation}
   -2a_{12}\ (36\ \times)\,.
\end{equation}
If the $SU(1,1)$ angles are chosen such
that $V_0$ is positive (de Sitter)
then the eigenvalues of the mass-matrix are also
positive for all 36 scalars.

In the present case $C_-<0$ in $Z_0$, which implies that
for the $SU(1,1)$ scalars we have a maximum. So in this example
there are two tachyons in the $SU(1,1)$ sector.
In \cite{dRWP1} we showed in the truncated model with two scalars that
$V_{11}$ and $V_{22}$, and therefore $C_-$,
change sign on a circle around $Z_0$. This
turns the maximum for the $SU(1,1)$ scalars into a minimum. It will be
interesting to see if this phenomenon also holds when all 36 matter
scalars are taken into account.

\subsection{$ SO(3,1)_-\otimes SO(3)_-\otimes  SO(3)_+$\label{G6}}

In this case the groups have to be arranged as follows:
\begin{equation}
R,S,\ldots =
\overbrace{\,1\ 2\ 3\ 10\ 11\ 12\,}^{i=1}\
\overbrace{\,4\ 5\ 6\,}^{i=2}\
\overbrace{\,7\ 8\ 9\,}^{i=3}\,.
\end{equation}
In this case the rotation subgroup of the  $SO(3,1)$ subgroup is
embedded on the indices $1\ldots 3$, the boosts on the indices
$10\ldots 12$.
Here we have:
\begin{eqnarray}
\label{C0min6}
  C_{0-} &=& -\half(g_1^2+g_2^2)\,,
    \\
 \label{T0min6}
  T_{0-}  &=& 2a_{12}\,,
    \\
\label{Del06}
  \Delta_0 &=& a_{12}^2\,,
    \\
\label{R0ii6}
  R^{(0\,ii)} &=&
   {1 \over |a_{12}|}\,\big(
    a_{i1}^2 + a_{i2}^2 \big)\,,\\
\label{V06}
  V_0 &=& -|a_{12}| -2a_{12}\,.
\end{eqnarray}

We have the following eigenvalues:
\begin{eqnarray}
  &&0\ (3\ \times),\ -2a_{12}\ (18\ \times),\
    2|a_{12}|-2a_{12}\ (9\ \times),\nonumber\\
  &&2|a_{12}|-4a_{12}\ (5\ \times),\
    2|a_{12}|+2a_{12}\ (1\ \times)\,.
\end{eqnarray}
For $a_{12}<0$ (de Sitter)
there are no negative eigenvalues, and four zero eigenvalues.
However, since $\sgn C_-<0$, the $SU(1,1)$
scalars produce the instability.

\subsection{$SO(3,1)_-\otimes SO(3,1)_-$\label{G7}}

In this case the groups have to be arranged as follows:
\begin{equation}
R,S,\ldots =
\overbrace{\,1\ 2\ 3\ 10\ 11\ 12\,}^{i=1}\
\overbrace{\,4\ 5\ 6\  7\  8\  9\,}^{i=2}\,.
\end{equation}
In this case the rotation subgroup of the two $SO(3,1)$ subgroups is
embedded on the indices $1\ldots 6$, the boosts on the indices
$7\ldots 12$.
Here $V_{0\,11}$ and $V_{0\,22}$ each are $-\half$, $W_{12}=1$,
and we have:
\begin{eqnarray}
\label{C0min7}
  C_{0-} &=& -\half(g_1^2+g_2^2)\,,
    \\
 \label{T0min7}
  T_{0-}  &=& 2a_{12}\,,
    \\
\label{Del07}
  \Delta_0 &=& a_{12}^2\,,
    \\
\label{R0ii7}
  R^{(0\,ii)} &=&
   {1 \over |a_{12}|}\,\big(
    a_{i1}^2 + a_{i2}^2 \big)\,,\\
\label{V07}
  V_0 &=& -|a_{12}| -2a_{12}\,.
\end{eqnarray}

To make $V_0$ positive we have to choose $a_{12}<0$.
The eigenvalues of the mass-matrix are:
\begin{eqnarray}
  &&0\ (6\ \times),\
    2|a_{12}|-2a_{12}\ (18\ \times),\nonumber\\
  &&2|a_{12}|-4a_{12}\ (10\ \times),\
    2|a_{12}|+2a_{12}\ (2\ \times)\,.
\end{eqnarray}
For $a_{12}<0$ there are no negative eigenvalues, and eight zero eigenvalues.
However, since $C_-<0$, the $SU(1,1)$ scalars produce the instability.

\subsection{${\rm SL}(3,\mathbb{R})_-\otimes {\rm SO}(3)_-$\label{G8}}

In this case the groups are arranged as follows:
\begin{eqnarray}
\stackrel{i=1}{\overbrace{\,1\ 2\ 3\ 4\ 5\ 7\ 8\ 9\,}}\
\stackrel{i=2}{\overbrace{\,4\ 5\ 6\,}}\,.
\end{eqnarray}
The rotation subgroup of $SL(3,\mathbb{R})$ is placed on the indices
$7,\ldots 9$, the noncompact generators on $1,\ldots 5$.
In this case we have
\begin{eqnarray}
\label{C0min8}
  C_{0-} &=& -\half(g_1^2+g_2^2)\,,
    \\
 \label{T0min8}
  T_{0-}  &=& 2a_{12}\,,
    \\
\label{Del08}
  \Delta_0 &=& a_{12}^2\,,
    \\
\label{R0ii8}
  R^{(0\,ii)} &=&
    {1 \over |a_{12}|}\,\big(
    a_{i1}^2 + a_{i2}^2 \big)\,,\\
\label{V08}
  V_0&=&-|a_{12}|-2a_{12}\,.
\end{eqnarray}

The eigenvalues are:
\begin{eqnarray}
   &&0\ (5\ \times),\ -2a_{12}\ (6\ \times),\
   6|a_{12}|+4a_{12}\ (3\ \times),\
   6|a_{12}|-2a_{12}\ (15\ \times),\ \nonumber\\
   &&6|a_{12}|-6a_{12}\ (7\ \times),\
\end{eqnarray}
In the de Sitter case ($a_{12}<0$) there are no negative eigenvalues, but
of course the $SU(1,1)$ scalars are tachyons because $C_-<0$. The five
zero eigenvalues are associated to the Goldstone fields
which give mass to the gauge fields corresponding to the
non--compact gauge generators.

In the AdS case ($a_{12}>0$) the potential in the extremum is
$V_0=-3|a_{12}|$,  in that case there are seven additional
zero eigenvalues.

\subsection{${\rm SU}(2,1)_+\otimes {\rm SO}(2,1)_+$\label{G9}}

The gauge generators are arranged as follows:
\begin{eqnarray}
\stackrel{i=1}{\overbrace{\,1\ 2\ 3\ 4\ 7\ 8\ 9\ 10\,}}\
\stackrel{i=2}{\overbrace{\,5\ 6\ 11\,}}\,.
\end{eqnarray}
The compact generators of $SU(2,1)$ are placed on the indices
$7,\ldots 10$,  the noncompact ones on $1,\ldots 4$.
In this case we have:
\begin{eqnarray}
\label{C0min9}
  C_{0-} &=& 6g_1^2 + \half g_2^2\,,
    \\
 \label{T0min9}
  T_{0-}  &=& 0 \,,
    \\
\label{Del09}
  \Delta_0 &=& 12 a_{12}^2\,,
    \\
\label{R0ii9}
  R^{(0\,ii)} &=&
   {1 \over \sqrt{3}|a_{12}|}\,\big(
    6a_{i1}^2 + \half\,a_{i2}^2 \big)\,,\\
\label{V09}
  V_0 &=& 2\,\sqrt{3}\, |a_{12}|\,>\,0\,.
\end{eqnarray}

Again we find a de Sitter vacuum.
The mass of the $SU(1,1)$ scalars
is $2\sqrt{3}|a_{12}|$
while in the matter sector the mass eigenvalues are:
\begin{eqnarray}
&& 0\ (6\ \times),\ \sqrt{3}|a_{12}|\ (12\ \times),\
2\sqrt{3}|a_{12}|\ (10\ \times),\ \sqrt{3}|a_{12}|(1\pm 2\sqrt{2}\ (4\ \times)\,.
\end{eqnarray}
The spectrum contains four tachyonic modes. The six zero modes correspond
to the Goldstone bosons of the broken noncompact generators.

\section{Conclusions\label{Conc}}

In this paper we have searched in gauged $N=4$
supergravity with semi-simple gauge groups  for examples that give a
positive cosmological constant and
a non-negative mass-matrix for all scalar fields. In all examples
considered the scalar potential does allow a positive extremum,
but we have always found tachyons. In our search we limited
ourselves to six vector multiplets, and therefore 36 matter scalars.

We found that there are two classes of gauge groups in the
nine that were considered. Five have a positive extremum for
all values of the parameters in the problem (coupling constants,
$SU(1,1)$ angles), have tachyonic modes in the matter sector,
and positive $m^2$ for the $SU(1,1)$ scalars. In four
cases we find that the sign of potential in the extremum
depends on the choice of parameters, that the matter scalars
all have positive $m^2$ (if the parameters are chosen such that
the extremum occurs for positive potential),
and the two $SU(1,1)$ scalars are the tachyons.
These last four are precisely all the cases with an $SO(3)_-^2$ subgroup.
This distinction is not yet understood.

Certain features of the mass spectrum are clear. We always find
the appropriate number of massless modes to provide for the
Goldstone bosons of the broken gauge symmetries. In a number
of cases we have more zero modes than Goldstone bosons.
This might be an indication that such models can be embedded
in a gauged supergravity with $N>4$, and that these
extra zero modes are related to Goldstone bosons that occur
in the larger supergravity theory.

Another feature of gauged supergravity theories, remarked in
\cite{KLPS}, is the fact that the mass spectrum of gauged
supergravity is often such that $3m^2/V_0$ is an integer.
This is an interesting observation, since it makes such models
unsuitable as candidates for slow-roll inflationary scenarios.
We find the same property for the ratio of mass and potential
in cases where the gauge group is a product of two simple groups,
with one exception in the $SU(2,1)_+\otimes SO(2,1)_+$ gauging
in Section \ref{G9}, where this ratio contains a factor $\sqrt{2}$.
For groups that have three simple factors and positive
cosmological constant, the ratio becomes
parameter dependent for some of the masses. However, also in these
cases one does not have enough freedom to tune the parameters such that
all tachyon masses become small.

As far as supersymmetry breaking is concerned the analysis of
\cite{MdRPW1} can be applied. For the groups in Sections
\ref{G1}-\ref{G4} and \ref{G9}
we always have $V_0>0$, and supersymmetry is completely
broken. In Sections \ref{G5}-\ref{G8} $V_0$ depends on the sign of $a_{12}$,
and one finds that for $V_0>0$ supersymmetry is completely broken, for
$V_0\le 0$ $N=4$ supersymmetry is preserved. The supersymmetry variations
of the fermions in these last three cases are proportional to
$g_1\Phi_{(1)} - ig_2\Phi_{(2)}$, for which
\begin{equation}
   |g_1\Phi_{(1)} - ig_2\Phi_{(2)}|^2 = R^{(0\,11)}+R^{(0\,22)}+2I^{(0\,12)}
      = 2\,(|a_{12}| - a_{12})\,.
\end{equation}
This vanishes for $a_{12}>0$, leading to unbroken supersymmetry in
AdS spacetime. In these cases the potential in the AdS extremum
is $V_0=-3|a_{12}|$. This value follows also from the integrability
condition arising from the supersymmetry variation of the gravitinos.

The relation of our $N=4$ work with the $N=2$ results of
\cite{FTVP} remains intriguing. There are three cases presented in
\cite{FTVP}. The one which seems most directly related to $N=4$
supergravity has five vector multiplets, gauging, with the
graviphoton, a six-dimensional $SO(2,1)\otimes SO(3)$ group. In
addition, the model has two hyper--multiplets, giving a total of
18 scalar fields. The scalar manifold is
\begin{equation}
\label{N=2manifold}
  \left[\,{SU(1,1) \over U(1)}\,\times\,{SO(2,4)\over SO(2)\times SO(4)}\,\right]
                      \times\,\left[\,{SO(4,2)\over SO(4)\times SO(2)}\,\right]\,,
\end{equation}
where the last factor corresponds to the hyper--multiplets. The
two $SU(1,1)$ scalars play a similar role as in $N=4$ and allow
the introduction of $SU(1,1)$ mixing angles in the coupling to the
vectors. The gauge group is embedded in both $SO(4,2)$ groups, and
it was shown in \cite{FTVP} how to obtain the manifold
(\ref{N=2manifold}) from a truncation of
\begin{equation}
\label{N=4manifold}
     {SU(1,1) \over U(1)}\,\times\,{SO(6,6)\over SO(6)\times SO(6)}\,.
\end{equation}
In the $N=4$ case the gauge group would be $[SO(2,1)\times
SO(3)]^2$, the $N=2$ group being the diagonal subgroup. The only
way to embed this group in $N=4$ is as $SO(3)_-\times
SO(2,1)_+\times SO(2,1)_-\times SO(3)_+$, which we have in Table
2 in Section \ref{firstder} but which does not have an extremum in
the matter and $SU(1,1)$ scalars. We may restrict our analysis to
points in the moduli space preserving the diagonal compact
subgroup $SO(2)_D\times SO(3)_D$ of the gauge group (in this
analysis we set the $g_1=g_4$, $g_2=g_3$, $\alpha_1=\alpha_4$,
$\alpha_2=\alpha_3$) . To this end it suffices to study the
behavior of the scalar potential as a function of the
$SO(2)_D\times SO(3)_D$ singlets only (indeed,  the scalar
potential being invariant in particular under $SO(2)_D\times SO(3)_D$,
its dependence on scalar fields which are not singlets with
respect to this group will be at least quadratic, and therefore in
order to analyse the critical points of the potential we may set
these fields to zero). These singlets in the matter sector are
four. This can be seen by first considering the action of the
compact subgroup $SO(3)_-\times SO(2)_+\times SO(2)_-\times
SO(3)_+$ under which the matter scalar fields transform as
follows:
\begin{eqnarray}
{\bf (3+1,1,2,1)}+{\bf (1,2,1,3+1)}+{\bf (1,2,2,1)}+{\bf (3+1,1,1,3+1)}\,.
\end{eqnarray}
The scalars in ${\bf (3+1,1,2,1)}$ and ${\bf (1,2,1,3+1)}$ correspond
to the 16 scalars in the \break $SO(2,4)/SO(2)\times SO(4)$ and
 $SO(4,2)/SO(4)\times SO(2)$ cosets,
those in ${\bf (1,2,2,1)}$ and \break ${\bf (3+1,1,1,3+1)}$ will be
projected out by the $N=4\rightarrow N=2$ truncation.
Branching the above representations
with respect to $SO(2)_D\times SO(3)_D$
we obtain two singlets $\varphi_{1,2}$
 from ${\bf (1,2,2,1)}$, one singlet
$\xi$ from ${\bf (3,1,1,3)}$ and finally the ${\bf (1,1,1,1)}$ singlet
 $\xi_0$. A non vanishing value
for $\xi$ would therefore break
$SO(3)_-\times SO(3)_+$ to $SO(3)_D$.
Although the potential does not have
critical points in the $SU(1,1)$ and
$\varphi_{1,2},\,\xi_0,\,\xi$ scalars,
a numerical analysis shows that it can
be extremized with respect to all the
scalars except $\xi$, for large enough values
of $|\xi|$ (at the origin of the
matter sector the potential does not have
an extremum  with respect to the
$SU(1,1)$ scalars). In other words there is
an infinite range of values for $\xi$
for which all the other scalars can
be fixed so that the potential has
a run--away behavior in $\xi$ field only.
In these points the potential is positive.
It seems therefore that in lifting the
$N=2$ model with vector and hyper--multiplets to
$N=4$ new scalar fields emerge which destabilize
the $N=2$ de Sitter vacuum.\par

%
%
The program presented in this paper can be extended in many
directions. One possibility could be to consider contractions
($CSO$ groups) of the gauge groups studied here.  Also the present
analysis can be generalized to include Peccei--Quinn symmetries.
These symmetries naturally appear in Scherk--Schwarz reductions
 \cite{SS,BdRE,ADFL},
therefore such an investigation could also elucidate the relation
between $N=4$ supergravity with nonzero $SU(1,1)$ angles and
string theory. Gaugings related to Scherk--Schwarz reductions from
$D=5$ could also provide new ways for obtaining some of the $N=2$
models with stable de Sitter vacua \cite{FTVP} as effective
realizations of a larger gauged $N=4$ theory (although a definite
statement about this possibility would require an analysis that we
postpone to future work). As an example we could consider the
$N=4$ supergravity coupled to six matter multiplets and gauge a
group of the form $G_{S-S}\times SO(2,1)\times SO(3)$ where
$G_{S-S}$ is a non--semi-simple gauge group \'a la Scherk--Schwarz.
It was shown indeed that, for a certain choice of the gauge
parameters, the effect of $G_{S-S}$ alone amounts to a partial
supersymmetry breaking from $N=4$ to $N=2$, in which the final
effective supergravity is coupled to five vector multiplets and no
hyper--multiplet. This is the ungauged version of one of the
models considered in \cite{FTVP}. Therefore if we gauge in the
$N=4$ theory the group $G_{S-S}\times SO(2,1)\times SO(3)$ and
introduce $SU(1,1)$ angles for each simple factor, we would be
left at the level of the $N=2$ effective model with a surviving
$SO(2,1)\times SO(3)$ gauge group. However, in order to recover in
this framework one of the models without hyper--multiplets
constructed in \cite{FTVP} a crucial ingredient would be the
presence of the Fayet--Iliopoulos term corresponding to the
$SO(3)$ factor, whose $N=4$ origin is still unclear.\par
 There are (at least)
two aspects of this program which remain to be elucidated. The first is the
existence or non-existence of stable de Sitter vacua in $N=4$ supergravity,
the second is the relation of gauged $N=4$ supergravity with $SU(1,1)$ angles
with ten and/or eleven dimensions. If such a relation could be
established, the no-go theorem of \cite{GWG,dWHDS,MN}  would come into play,
and one would know that to solve the first problem would require
flux reduction or hyperbolic reduction and/or other ways around
the no-go theorem \cite{PT}.

\acknowledgments

\bigskip

MT would like to thank S. Ferrara for useful remarks.
The work of DBW is part of the research
programme of the ``Stichting
voor Fundamenteel Onderzoek van de Materie'' (FOM).
This work is supported in part by the European
Commission RTN programme HPRN-CT-2000-00131, in which MdR and DBW
are associated to the University of Utrecht. The work
of MT is supported by a European Community Marie Curie Fellowship
under contract HPRN-CT-2001-01276. SP thanks the ICTP in Trieste
and the Centre for Theoretical Physics in Groningen for their
hospitality.

\appendix

\section{$SU(1,1)$ scalars and angles\label{SU11}}

We parametrise the scalars of the $SU(1,1)/U(1)$ coset in a suitable
$U(1)$ gauge as
\begin{equation}
   \phi_1 = {1\over \sqrt{1-r^2}}\,,\qquad
   \phi_2 = {re^{i\varphi}\over \sqrt{1-r^2}}\,.
\end{equation}
The scalars $r$ and $\varphi$ then appear in the potential (\ref{pot}) through
\begin{eqnarray}
  R^{(ij)} &=& {g_i g_j\over 2}  (\Phi^*_i\Phi_j + \Phi^*_j\Phi_i)\nonumber\\
           &=& g_ig_j\left(\cos(\alpha_i-\alpha_j){1+r^2\over 1-r^2}
                - {2r\over 1-r^2}\cos(\alpha_i+\alpha_j+\varphi)
                \right)\,,\\
  I^{(ij)} &=& {g_ig_j \over 2i} (\Phi^*_i\Phi_j - \Phi^*_j\Phi_i)\nonumber\\
           &=& -g_ig_j\sin(\alpha_i-\alpha_j)\,.
\end{eqnarray}
Introducing
\begin{eqnarray}
\label{CS}
  C_{\pm} &=& \sum_{ij} g_ig_j\cos(\alpha_i\pm\alpha_j)V_{ij}\,,\quad
  S_{+} = \sum_{ij} g_ig_j\sin(\alpha_i+\alpha_j)V_{ij} \,,\\
  T_{-}   &=& \sum_{ij} g_ig_j\sin(\alpha_i-\alpha_j)W_{ij} \,,
\end{eqnarray}
we rewrite the potential as
\begin{equation}
\label{potrphi}
   V = C_-\,{1+r^2\over 1-r^2} - {2r\over 1-r^2}\,
     \big(C_+\cos\varphi - S_+\sin\varphi\big) - T_-\,.
\end{equation}
This extremum in $r$ and $\varphi$ takes on the form
\begin{eqnarray}
  \cos\varphi_0 &=& {s_1  C_+\over \sqrt{C_+^2 + S_+^2}}\,,\quad
  \sin\varphi_0 = -{s_1 S_+\over \sqrt{C_+^2 + S_+^2}}\,,
  \nonumber\\
    r_0 &=& {1\over \sqrt{C_+^2 + S_+^2}}
    \left( s_1 C_- + s_2 \sqrt{\Delta}\right)\,,\qquad
    \Delta\equiv C_-^2-C_+^2-S_+^2\,,
\end{eqnarray}
where $s_1$ and $s_2$ are signs. These are determined by
requiring $r_0\ge 0$ and $r_0<1$, this gives $s_1=\sgn C_-$ and
$s_2=-1$.
Substitution of $r_0$ and $\varphi_0$ in $V$ leads to
\begin{equation}
  V_0 = \sgn{C_-}\,\sqrt{\Delta}  - T_- \,.
\end{equation}

In the case that all $SU(1,1)$ angles $\alpha_i$
vanish,  $S_+=T_-=0$ and $C_-=C_+$, and one finds
$r_0=1$ and $\Delta=0$. This is a singular point of the
parametrisation, which we will exclude.  It is
generalisation of the Freedman-Schwarz potential \cite{FS}
to the case of general matter coupling.

For the kinetic term and mass-matrix of the $SU(1,1)$ scalars we
introduce:
\begin{eqnarray}
x' &=& {2\over (1-r_0)^2}(r\cos\varphi - r_0\cos\varphi_0)\,,
 \nonumber\\
y' &=& {2\over (1-r_0)^2}(r\sin\varphi - r_0\sin\varphi_0)\,.
\end{eqnarray}
In these variables we find
\begin{eqnarray}
\label{actionphi}
   {\cal L}_{\phi} &=&
   - \half \left( {1-r_0^2\over 1-r^2}\right)^2
    \big( \partial_\mu x'\partial^\mu x' + \partial_\mu y' \partial^\mu y'\big)
   - V_0
   -\half\, \sgn{C_-}\,\sqrt{\Delta}\,\big( x^{\prime\,2}
    + y^{\prime\,2}\big)\ +\ \ldots\,.
\end{eqnarray}
It is clear that we have two tachyons for $\sgn C_-<0$, and
two positive mass scalars for $\sgn C_->0$.

It is useful to also analyse the kinetic term of the vectors,
since positivity\footnote{In \cite{dRWP1} it was
incorrectly stated that the kinetic terms of the vectors
acquire the wrong sign. The discussion below should clarify
this point.} of these terms might give further constraints
on the $SU(1,1)$ scalars, to which the vectors couple.
In $Z=Z_0$ we have
\begin{eqnarray}
  {\cal L}_{\rm kin, A}
  &=& -\quart F_{\mu\nu}{}^{+A}  F^{\mu\nu\,+A}\,
    \bigg( -{\phi^1{}_{(A)} - \phi^2{}_{(A)}\over \Phi_{(A)}}
       + {2\over \big|\Phi_{(A)}\big|^2} \bigg)
   \nonumber\\
  &&\qquad\qquad
     -\quart F_{\mu\nu}{}^{+I}  F^{\mu\nu\,+I}\,
    \bigg({\phi^1{}_{(I)} - \phi^2{}_{(I)}\over \Phi_{(I)}}\bigg)  + {\rm h.c}\,.
\end{eqnarray}
Here $\Phi_{(R)}\equiv e^{i\alpha_R}\phi^1 + e^{-i\alpha_R}\phi^2$.
We find after some algebra:
\begin{equation}
    -\quart F_{\mu\nu}{}^{+A}  F^{\mu\nu\,+A} S^{(A)}
    -\quart F_{\mu\nu}{}^{+I}  F^{\mu\nu\,+I} S^{*\,(I)} + {\rm h.c.}\,.
\end{equation}
Here $S^{(R)}$ is given by
\begin{equation}
  S^{(R)} = {1+re^{i(\varphi + 2\alpha_R)}\over
             1-re^{i(\varphi + 2\alpha_R)}}\,.
\end{equation}
The imaginary part of $S$ gives a total derivative in the kinetic term.
Therefore the kinetic terms are determined by
\begin{equation}
   Re S^{(R)} = {1-r^2\over 1 + r^2 - 2r\cos(\varphi+2\alpha_R)}\,,
\end{equation}
showing that the domain of positivity is $r<1$. In the extremum for the
$SU(1,1)$ scalars, and in the origin of the matter scalar manifold,
this becomes for the i'th factor of the gauge group
\begin{equation}
   Re S^{(i)}  = {g_i^2\over R^{(0\,ii)}}\,,
\end{equation}
where $R^{0\,ii}$ is given in (\ref{R0ii}). Indeed, we find that
always  $R^{0\,ii}>0$.

\section{Generators and structure constants\label{SCS}}

Our conventions for the structure constant are the following:
\begin{eqnarray}
     [T_R,\,T_S] = if_{RS}{}^U T_U\,,\quad f_{RST}\equiv f_{RS}{}^U\eta_{TU}\,,
\end{eqnarray}
where the structure constants $f$ are real.
The structure constants
$f_{RST}$ are completely antisymmetric, since $\eta{RS}$ is proportional to
the Cartan-Killing metric $\tr T_RT_S$ of the gauge group. The factor 2 in this
proportionality we absorb in a redefinition of the coupling constants $g_i$.
Our choice for the generators is based on the Gell-Mann matrices
(extended to $4\times 4$ matrices to treat $SO(3,1)$ in the same context):
\begin{eqnarray}
  &&\lambda_1 = \left(\begin{array}{cccc}
                     0 & 1 & 0 & 0\cr
                     1 & 0 & 0 & 0\cr
                     0 & 0 & 0 & 0\cr
                     0 & 0 & 0 & 0
                     \end{array}\right)\,,\quad
    \lambda_2 = \left(\begin{array}{cccc}
                     0 &-i & 0 & 0\cr
                     i & 0 & 0 & 0\cr
                     0 & 0 & 0 & 0\cr
                     0 & 0 & 0 & 0
                     \end{array}\right)\,,\quad
    \lambda_3 = \left(\begin{array}{cccc}
                     1 & 0 & 0 & 0\cr
                     0 &-1 & 0 & 0\cr
                     0 & 0 & 0 & 0\cr
                     0 & 0 & 0 & 0
                     \end{array}\right)\,,\ \nonumber\\
  &&\lambda_4 = \left(\begin{array}{cccc}
                     0 & 0 & 1 & 0\cr
                     0 & 0 & 0 & 0\cr
                     1 & 0 & 0 & 0\cr
                     0 & 0 & 0 & 0
                     \end{array}\right)\,,\quad
    \lambda_5 = \left(\begin{array}{cccc}
                     0 & 0 &-i & 0\cr
                     0 & 0 & 0 & 0\cr
                     i & 0 & 0 & 0\cr
                     0 & 0 & 0 & 0
                     \end{array}\right)\,,\quad
    \lambda_6 = \left(\begin{array}{cccc}
                     0 & 0 & 0 & 0\cr
                     0 & 0 & 1 & 0\cr
                     0 & 1 & 0 & 0\cr
                     0 & 0 & 0 & 0
                     \end{array}\right)\,,\ \nonumber\\
  &&\lambda_7 = \left(\begin{array}{cccc}
                     0 & 0 & 0 & 0\cr
                     0 & 0 &-i & 0\cr
                     0 & i & 0 & 0\cr
                     0 & 0 & 0 & 0
                     \end{array}\right)\,,\quad
    \lambda_8 = {1\over \sqrt{3}}\left(\begin{array}{cccc}
                     1 & 0 & 0 & 0\cr
                     0 & 1 & 0 & 0\cr
                     0 & 0 &-2 & 0\cr
                     0 & 0 & 0 & 0
                     \end{array}\right)\,,\ \nonumber\\
  &&\lambda_9 = \left(\begin{array}{cccc}
                     0 & 0 & 0 & 1\cr
                     0 & 0 & 0 & 0\cr
                     0 & 0 & 0 & 0\cr
                     1 & 0 & 0 & 0
                     \end{array}\right)\,,\quad
     \lambda_{10} = \left(\begin{array}{cccc}
                     0 & 0 & 0 & 0\cr
                     0 & 0 & 0 & 1\cr
                     0 & 0 & 0 & 0\cr
                     0 & 1 & 0 & 0
                     \end{array}\right)\,,\quad
     \lambda_{11} = \left(\begin{array}{cccc}
                     0 & 0 & 0 & 0\cr
                     0 & 0 & 0 & 0\cr
                     0 & 0 & 0 & 1\cr
                     0 & 0 & 1 & 0
                     \end{array}\right)\,.
\end{eqnarray}
The groups we consider have compact generators, which in the following we
will denote by $L_R$, and noncompact generators, denoted by $K_R$. We
will specify the structure constants with three $L$'s, and with
two $K$'s and one $L$.
The choice of the generators
of the simple groups used in this paper are summarized in Table~4.
\begin{center}
\renewcommand{\arraystretch}{1.5}
\begin{tabular}{|l||l||}
\hline
Group\hfil    & Generators (L;\ K)        \\
\hline
$SO(3)$       & $L_\alpha = \lambda_7,-\lambda_5,\lambda_2$
         \\
$SO(2,1)$     & $L_3 = \lambda_2,\ K_{1,2}=i\lambda_6,i\lambda_4,$
         \\
$SO(3,1)$     & $L_\alpha=\lambda_7,-\lambda_5,\lambda_2,\
                 K_\alpha=
                 i\lambda_9,i\lambda_{10},i\lambda_{11}$
        \\
$SL(3,\mathbb{R})$ & $L_\alpha =\lambda_7,-\lambda_5,\lambda_2,\
                  K_\alpha=i\lambda_{6},i\lambda_{4},i\lambda_1,\
                  K_4=i\lambda_3,K_5=i\lambda_8$
         \\
$SU(2,1)$     & $L_\alpha=\lambda_1,\lambda_2,\lambda_3,\ L_4=\lambda_8;\
                 K_1=i\lambda_4,\ K_2=i\lambda_{5},\ K_3=i\lambda_{6},\
                 K_4=i\lambda_7$
        \\
\hline
\end{tabular}
\renewcommand{\arraystretch}{1.0}
\end{center}
\vspace{.25truecm}
\noindent {\bf Table\ 4.} \ \
Generators for the simple groups used in this paper. We always have
$\alpha,\beta,\gamma=1,\ldots 3$. \\

The structure constants for the {\bf $SO(3)$} subgroups are in all cases
\begin{equation}
   [L_\alpha,\ L_\beta] =
   i\epsilon_{\alpha\beta\gamma} L_\gamma\,,\ (\alpha,\beta,\gamma= 1,2,3)\,.
\end{equation}
For {\bf $SO(2,1)$} we have
\begin{equation}
   [L_3,K_\beta] =
   -i\epsilon_{3\beta\gamma}K_\gamma\,,\ (\beta,\gamma=1,2)\,.
\end{equation}
In the case of {\bf $SO(3,1)$} the commutation relation involving noncompact
generators are
\begin{equation}
   [L_\alpha,K_\beta] = i\epsilon_{\alpha\beta\gamma}K_\gamma\,,\
   (\alpha,\beta,\gamma= 1,2,3)\,.
\end{equation}
For {\bf $SL(3,\mathbb{R})$} the relations between $L$ and $K$ are:
\begin{eqnarray}
   &&[L_\alpha,K_\beta] = -i\epsilon_{\alpha\beta\gamma}K_\gamma\,,\
   (\alpha\ne\beta,\ \alpha,\beta,\gamma= 1,2,3)\,, \nonumber\\
   &&[L_1,K_1] = i(K_4-\sqrt{3}K_5)\,,\ [L_2,K_2] = i(K_4+\sqrt{3}K_5)\,,\
   [L_3,K_3] = -2iK_4\,.
\end{eqnarray}
Finally, for {\bf $SU(2,1)$} we have
\begin{eqnarray}
      [L_\alpha,\ L_\beta] =
   2i\epsilon_{\alpha\beta\gamma} L_\gamma\,,\ (\alpha,\beta,\gamma= 1,2,3)\,,\
      [L_\alpha,L_4]=0\,.
\end{eqnarray}
The commutation relation between the remaining generators are given
in Table 5. The structure constants
presented for the different groups are the $f_{RS}{}^T$. For use in
the calculation of second derivatives etc. the index $T$ has to be
lowered by $\eta_{TU}$, which may give a sign depending on the embedding
of the group in $SO(6,6)$. Other commutation relations, such as
$[K_\alpha,\ K_\beta]$ follow from the antisymmetry of $f_{RST}$.

\begin{center}
\renewcommand{\arraystretch}{1.5}
\begin{tabular}{|c|c|c|c|c|}
\hline
\hfil    & $K_1=i\lambda_4$  & $K_2=i\lambda_6$
         & $K_3=i\lambda_5$ &  $K_4=i\lambda_7$ \\
\hline
  $L_1=\lambda_1$ & $ K_4$ & $ K_3$ & $-K_2$ & $-K_1$ \\
  $L_2=\lambda_2$ & $ K_2$ & $-K_1$ & $K_4 $ & $-K_3$ \\
  $L_3=\lambda_3$ & $ K_3$ & $-K_4$ & $-K_1$ & $K_2 $ \\
  $L_4=\lambda_8$ & $\sqrt{3}K_3$ &
  $\sqrt{3}K_4$ & $-\sqrt{3}K_1$ & $-\sqrt{3}K_2$ \\
\hline
\end{tabular}
\renewcommand{\arraystretch}{1.0}
\end{center}
\vspace{.25truecm}
\noindent {\bf Table\ 5.} \ \
Commutation relations between compact and noncompact generators in $SU(2,1)$.
The table reads $[L_1,K_1]= iK_4$, etc. \\

\section{The parameters $P$\label{PGB}}

The independent scalars are contained in $G$ and $B$. Define
$G_\pm \equiv (G\pm B)$, and $P\equiv G_+$.
Then $G=\half(P+P^T),\ B=\half(P-P^T)$, and we find
\begin{eqnarray}
{\partial G_{cd}\over \partial P_{ab}} &=&
       \half(\delta_{ac}\delta_{bd}+\delta_{ad}\delta_{bc})\,,
\quad
{\partial B_{cd}\over \partial P_{ab}} =
       \half(\delta_{ac}\delta_{bd}-\delta_{ad}\delta_{bc})\,,
\nonumber\\
{\partial G^{-1}{}_{cd}\over \partial P_{ab}} &=&
  - \half\,\bigg(G^{-1}{}_{ca}G^{-1}{}_{bd}+G^{-1}{}_{cb}G^{-1}_{ad}\bigg)\,.
\end{eqnarray}
In the study of the potential we need the first and second derivatives
of $X$ and $Y$ with respect to $P$. We find:
\begin{eqnarray}
{\partial X_{cd}\over \partial P_{ab}}\bigg|_0 &=& 0\,,\nonumber\\
{\partial Y_{cd}\over \partial P_{ab}}\bigg|_0 &=& \delta_{ad}\delta_{bc}\,,\nonumber\\
{\partial^2 X_{ef}\over \partial P_{ab}\partial P_{cd}}\bigg|_0 &=&
   \half\delta_{ac}\,(\delta_{de}\delta_{bf}+ \delta_{be}\delta_{df})\,,\nonumber\\
{\partial^2 Y_{ef}\over \partial P_{ab}\partial P_{cd}}\bigg|_0 &=&
   -\half(\delta_{bc}\delta_{de}\delta_{af}
   + \delta_{ad}\delta_{be}\delta_{cf})\,,\nonumber\\
\end{eqnarray}

\end{document}